\documentclass{webofc}
\usepackage[varg]{txfonts}   
\usepackage{hyperref}

\begin{document}
\title{Improving the muon track reconstruction of IceCube and IceCube-Gen2}
\author{\firstname{Federica} \lastname{Bradascio}\inst{1}\fnsep\thanks{\email{federica.bradascio@desy.de}} \and
        \firstname{Thorsten} \lastname{Gl\"usenkamp}
        \inst{2}\fnsep\thanks{\email{thorsten.gluesenkamp@fau.de}}
}

\institute{DESY, D-15735 Zeuthen, Germany
\and
           Erlangen Centre for Astroparticle Physics, Friedrich-Alexander-Universit\"at Erlangen-N\"urnberg, D-91058 Erlangen, Germany
        }

\abstract{%
  IceCube is a cubic-kilometer Cherenkov telescope operating at the South Pole. Its goal is to detect astrophysical neutrinos and identify their sources. High-energy muon neutrinos are identified through the secondary muons produced via charge current interactions with the ice. The present best-performing directional reconstruction of the muon track is a maximum likelihood method which uses the arrival time distribution of Cherenkov photons registered by the experiment's photomultipliers. Known systematic shortcomings of this method are to assume continuous energy loss along the muon track, and to neglect photomultiplier-related effects such as prepulses and afterpulses. This work discusses an improvement of about $20 \%$ to the muon angular resolution of IceCube and its planned extension, IceCube-Gen2. In the reconstruction scheme presented here, the expected arrival time distribution is now parametrized by a predetermined stochastic muon energy loss pattern. The inclusion of pre- and afterpulses modelling in the PDF has also been studied, but no noticeable improvement was found, in particular in comparison to the modification of the energy loss profile.
}
\maketitle
%

\section{Introduction}
\label{sec-1}
IceCube is a cubic-kilometer scale Cherenkov telescope operating at the South Pole \cite{ic}, detecting neutrinos of all flavors with energies between tens of GeV and several PeV. IceCube discovered high-energy astrophysical neutrinos in 2013 \cite{discovery}, and has since found compelling evidence of a potential high-energy neutrino source \cite{txs}. 
IceCube-Gen2 is a planned next generation neutrino observatory which will embed the current IceCube array, adding 120 new strings with a larger spacing \cite{gen2}. It will increase the instrumented volume by a factor of 10 and will be able to detect sources $\sim$5 times fainter.

IceCube consists of 5160 digital optical modules (DOMs), each containing a 10-inch photomultiplier tube (PMT). The PMTs detect Cherenkov light emitted from charged secondary particles created in neutrino interactions. High-energy neutrinos are thus identified through two main channels: tracks and cascades. Tracks are produced by charged-current interactions of $\nu_{\mu}$ and cascades arise from relatively short particle showers induced by charged-current interactions of $\nu_{e}$ and $\nu_{\tau}$, as well as by neutral-current interactions of all neutrino flavors. 
The muon tracks are well-suited for point-source searches, since they have an angular resolution of $<2^{\circ}$. Their exact resolution varies, depending on their energy. Improving the pointing accuracy for these tracks therefore directly translates into a better discovery potential of the sources of astrophysical neutrinos. 

In this paper, the main limiting factors of the current muon track reconstruction will be analyzed and a modification will be presented that can partially overcome them. The resulting new reconstruction improves the old one by about $20 \%$.

\section{Muon track reconstruction in IceCube}
\label{sec-2}
The angular reconstruction of muon tracks in IceCube relies on the optical detection of the Cherenkov radiation emitted by the muon interactions in the ice and registered by the PMTs \cite{trackreco}. 
The arrival time of the photons is defined as the time residual, $t_{\mathrm{res}}$, which is the difference between the detection time and the expected arrival time of a photon that does not scatter. 

The present muon reconstruction used in IceCube is called \emph{SplineMPE}, and is based on a maximum likelihood method assuming a minimum ionizing energy loss of an infinite muon track \cite{splinempe}.
It evaluates the arrival time probability density function (PDF) of the first hit of N detected photons, since it is less scattered and thus carries more directional information than photons arriving later. 
The total likelihood is given by:
\begin{equation}
\label{eq:mpe}
    \mathcal{L}_{\mathrm{MPE}}= \prod_{i=1}^{N_{\mathrm{DOM}}} N_i \cdot p_i(t_{\mathrm{res},1})(1 - P_i(t_{\mathrm{res},1})^{(N-1)}),
\end{equation}
where $p_i(t_{\mathrm{res},1})$ and $P_i(t_{\mathrm{res},1})$ are the respective time residual PDF and its cumulative density function (or CDF) of the first photon for DOM $i$, $N_i$ is the number of observed hits for the given DOM, and $N_{\mathrm{DOM}}$ is the number of hit DOMs. The likelihood is then maximized varying the track parameters.
This is called \emph{Multi Photo Electron} (MPE) likelihood, because it takes into account all light arriving at a DOM, but is adjusted to only use the timing information of the first photon.  It is called ``\emph{Spline}''  MPE, because the PDFs are modeled by multidimensional splines which have been fitted to tabulated photon hits for a given Cherenkov emission hypothesis \cite{spline}. In this case the emission hypothesis is a minimum ionizing muon track.

\section{Improving the muon track reconstruction}
\label{sec-3}

The performance of the likelihood reconstruction method described in the previous section depends heavily on the quality of the underlying PDFs. The probability distribution of photon arrival times at a DOM for a certain muon track has to be as close to reality as possible. The effect of a correct PDF on the angular resolution can be studied by performing a simulation that uses the same hypothesis as in the reconstruction. In this simulation, the photon arrival time distribution matches exactly the one used for the spline PDF: the time of each detected photon is drawn from the PDF used in the reconstruction. This represents an ideal simulation where the systematic deviations between reconstruction and hypothesis are removed. 
This simulation has been performed for IceCube-Gen2 and shows that a correct description of the expected arrival time distribution of photons would increase the resolution by $\sim 40\%$ at 100~TeV and $\sim 45\%$  at 1~PeV demonstrating the importance of a correct description of the time residual distributions.

The photon arrival time probabilities can be improved in two ways. The first one is to model PMT-related effects, that are not accounted for in the spline PDFs. A detailed study has been performed to understand the impact of these effects, such as PMT noise, current saturation, prepulses, late pulses and afterpulses. However, none of these are a limiting factor for the reconstruction and the inclusion of these effects seems to have a negligible impact on the angular resolution.

The second way is to parameterize muon stochastic energy losses. The current \emph{SplineMPE} reconstruction only describes the Cherenkov light distribution of a muon which looses energy continuously. However, starting above $\sim1$~TeV, muons mostly loose energy stochastically via bremsstrahlung, pair production, and nuclear interactions. The effect of these processes is a production of shower-like depositions on top of the track signature. Light created in such stochastic losses has a different emission spectrum, and it influences the time residual distribution. Therefore, stochastic energy losses need to be included in the track reconstruction.

\section{The SegmentedSplineMPE reconstruction} 
\label{sec-4}

A new reconstruction, called \emph{SegmentedSplineMPE}, has been developed to include the effect of muon stochastic energy losses. It is a maximum likelihood reconstruction that builds on \emph{SplineMPE}, but that uses a cascade-segmented track hypothesis. The total hypothesis is composed of the cascades produced by stochastic energy losses, a constant DOM-dependent noise term, and potentially the infinite muon track hypothesis that is already implemented in \emph{SplineMPE}. The reconstruction steps are shown below.
\begin{enumerate}
    \item The \emph{SplineMPE} reconstruction is performed giving a track as output.
    \item The \emph{SplineMPE} track is used as a seed for an additional energy loss pattern reconstruction called \emph{Millipede} described in \cite{enreco}. The outcome of this reconstruction is a series of $n$ cascades of energy $E_j$ representing the stochastic energy losses fixed along the muon track.
    \item The total PDF at each DOM $i$ is given by the weighted sum of $n$ PDFs assuming each reconstructed cascade as a source of photon emission:
    \begin{equation}
        p_1(t_{\mathrm{res},i}) = \sum_{j=0}^{n} w_{j}p_{1,j}(t_{\mathrm{res},i}),
    \end{equation}
    where $w_{j} = \frac{\lambda_j}{\sum_{k=1}^n \lambda_k}$. The parameter $\lambda_j$ denotes the expected number of photons of source $j$ in a given DOM, and it is calculated for cascades, a constant noise contribution, and potentially a minimum ionizing muon. This PDF is then inserted in \autoref{eq:mpe} and the likelihood is maximized. 
\end{enumerate}

This reconstruction has been implemented in C++ within the IceCube software framework. As an independent project w.r.t. SegmentedSplineMPE, a  a Python-based version has also been developed, which we denote by \emph{UberMillipede}\footnote{The name derives from the segmented energy reconstruction described in \cite{enreco}.} in the following.
\emph{UberMillipede} uses the exact same model and likelihood, but also allows one to fit the energies in every iteration of track-positional parameters in an inner loop, if desired. It has been found, however, that such a profile likelihood fit does not converge to a good solution and performs worse due to fluctuations in the inner-loop fit outcome. If the energies are fixed from the beginning, \emph{UberMillipede} gives identical results to \emph{SegmentedSplineMPE}, as expected. Fixing the energy losses is therefore a key factor in the convergence to a good solution. 

\section{Results and conclusions}
\label{subsec-4}

\begin{figure*}
    \centering
    \includegraphics[height=4cm,clip]{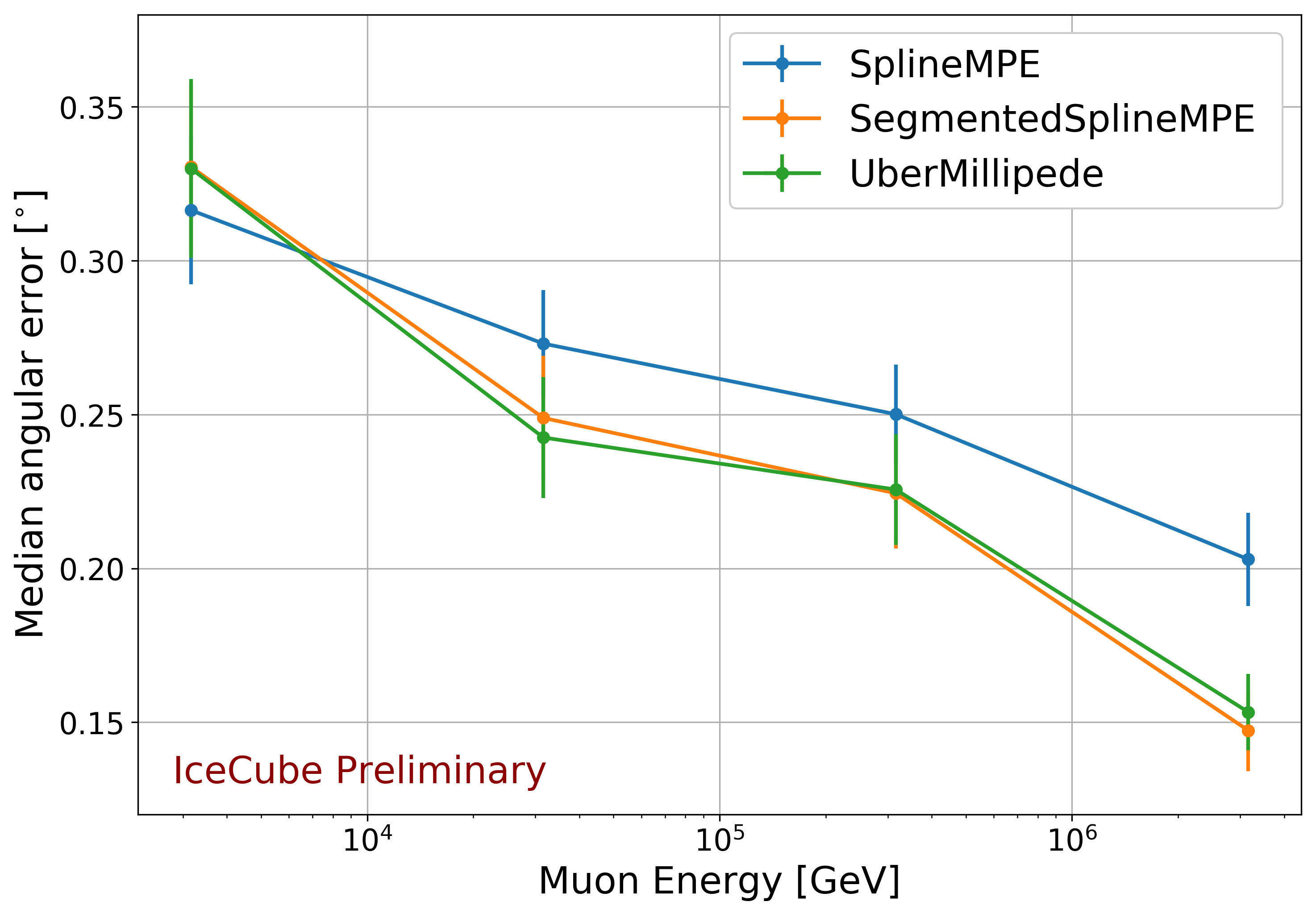}
    \includegraphics[height=4cm,clip]{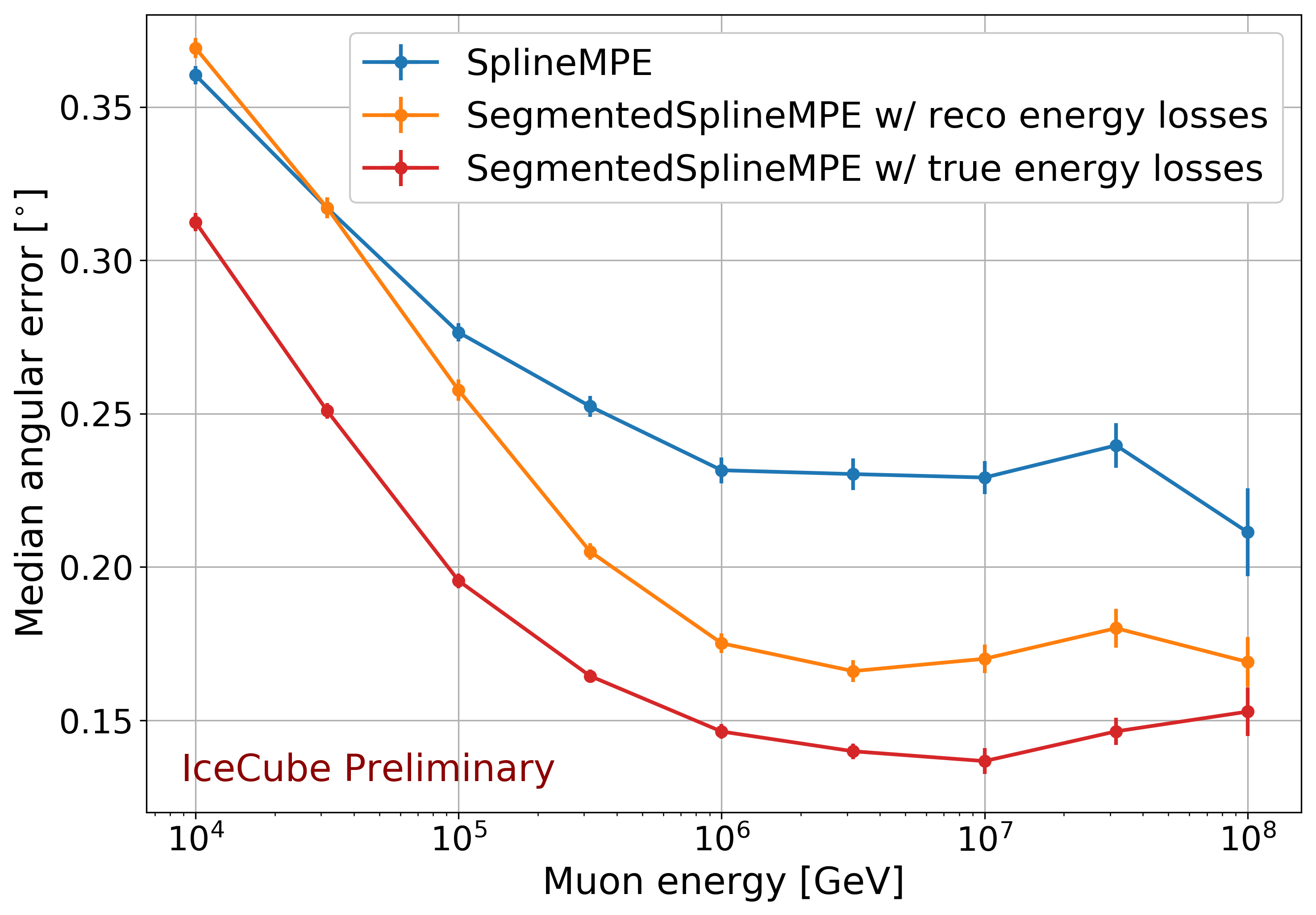} 
    \caption{Median angular resolution for IceCube (left) and IceCube-Gen2 (right) all-sky simulations. For IceCube, the \emph{SegmentedSplineMPE} reconstruction is compared to \emph{SplineMPE} and to \emph{UberMillipede}. 
    For IceCube-Gen2, the \emph{SegmentedSplineMPE} result seeded with true energy losses is also shown. The statistical error on the medians are calculated using bootstrapping technique.}
    \label{fig:ssmpe}       
\end{figure*}

The new reconstruction has been applied both to IceCube and IceCube-Gen2 all-sky simulations. \autoref{fig:ssmpe} shows the median angular error of \emph{SegmentedSplineMPE} compared to \emph{SplineMPE}. For IceCube, the results obtained with \emph{UberMillipede} are also shown for comparison: the two versions are compatible within the error bars. For IceCube-Gen2, the \emph{SegmentedSplineMPE} reconstruction using actual energy losses taken from the simulation output is compared to the one seeded with \emph{Millipede}-reconstructed cascades. The better resolution with true energy losses, especially at lower energies, shows that further improvements can be achieved if the cascade energy reconstruction can be improved.

\emph{SegmentedSplineMPE} improves the angular resolution of IceCube by $10\%$ at 100~TeV and $20\%$ at 1~PeV, and of IceCube-Gen2 by $10\%$ at 100~TeV and $25\%$ at 1~PeV. A key ingredient for convergence in the new reconstruction is the fixation of the energies in the beginning, instead of allowing them to vary in the inner loop of a profile-likelihood fit. For IceCube-Gen2, this resolution improvement has been translated in terms of integrated sensitivities and discovery potentials for an E$^{-2}$ astrophysical neutrino flux from a single source, using only horizontal tracks. After 15 years of operating IceCube-Gen2, the new reconstruction will increase the sensitivity to an astrophysical source at $90\%$ confidence level by $5\%$ with respect to the \emph{SplineMPE}. The IceCube-Gen2 studies are preliminary, and might change with more sophisticated IceCube-Gen2 simulations in the future. We also assume that we can reliably estimate the uncertainty on the new reconstruction, which is currently work in progress. 

\end{document}